# Information and Communication Technology in Migration: A Framework for Applications, Customization, and Research

Ali Arya, Luciara Nardon, Md Riyadh
Carleton University

**Abstract**

This paper addresses the role of Information and Communication Technology (ICT) in migration governance, support, and experience with particular attention to emerging technologies such as artificial intelligence, social media, and virtual reality. We propose a framework for technology use based on user groups and process types. We provide examples of using emerging technologies for migration-related tasks within the context of this framework. We then identify how such technologies can be applied to migration-related tasks, developed for customized use, and improved through research to add new features that can help different migration stakeholders. We suggest a series of possible directions for future research and development to take advantage of specific affordances of those emerging technologies more effectively.

## 1. Introduction

Migration, as "the movement of persons away from their place of usual residence, either across an international border or within a state" (IOM, 2019), has been an inseparable part of human experience (Tattersall, 2009). Migration is "part of a broader process of development and social transformation" (De Haas et al., 2019). So, an interdisciplinary combination of theories and insights is needed to understand and manage it. For example, labor market and social development theories are needed to explain the migration process and respond to it properly (De Haas et al., 2019). On the other hand, the development and use of technology have also been an essential characteristic of the human species, closely related to our evolution (Haidle, 2010). Technology has a ubiquitous presence in everyday life and is used by multiple stakeholders following different agendas from supporting to preventing migration, from bridges to walls. Information and Communication Technologies (ICT) are particularly important due to their widespread use and evolution in recent years. Understanding ICT and its related theories can play an essential role in migration studies. Migrants use these ICT tools to make decisions and execute their plans (Fortunati et al., 2013). Governments, organizations, and communities also use technology to understand and manage the migration process (Buono et al., 2019; Cukier, 2016). In turn, migration provides resources and demands that significantly influence the development of new technologies (Hübler, 2016; Ganguli et al., 2020).

The COVID-19 pandemic has affected the migration process drastically due to the movement restrictions, but at the same time, it emphasized the role of ICT in not only virtual communication but also virtual workplaces (Kniffin et al., 2021). While the health-related COVID restrictions are slowly removing, it is expected that many of the effects will be long-lasting. These effects strengthen the connection between migration and ICT with the possibility of increased use of technology for remote work, data collection, and communication, among other purposes.

The dynamic connections between technology and migration have been the subject of many scholarly studies (Pannocchia et al., 2020; Buono et al., 2019; Cukier, 2016; Nardon et al., 2015). Despite these efforts, an interdisciplinary framework that relates the affordances of emerging ICT to various migration stakeholders' needs is still missing. Such a framework is required to guide the effective use of ICT, develop customized ICT tools, and research new ICT features to help in various areas of migration.

The emergence of multiple ICT technologies such as social media, digital storytelling, computer games, data visualization, artificial intelligence (AI), virtual reality (VR), and mobile devices have the potential to change the dynamics of migration and migration-related decision making. Despite its ubiquity and promise, ICT is a complex tool; its affordances may not be clear to stakeholders in different fields; and it can be potentially dangerous if improperly or maliciously used. Thus, it is critical to understand the impact of ICT on the migration process, how, where, and when to use it, regulate it, and improve it. This understanding starts with the basic ICT processes (what types of operations ICT can do) and common and emerging ICT tools (specific technologies) and associates those with basic migration-related tasks. The insights generated through such associations can help those aiming to use ICT in migration tasks to know what is possible and what potentials exist. For example, knowing that generating novel data is an essential ICT affordance, and that crowdsourcing through social media is an emerging tool to do it, can suggest potentials for collecting information from migrants (at the government level), educating them (at the support organizations level), or supporting their decision-making (at the individual level). A framework that defines those affordances and tools and establishes such associations can be a starting point for ideation when planning to use ICT in migration tasks.

Despite the burgeoning literature on technology and migration, most studies to date are either focused on specific cases of technology use or cover the perspective of one discipline. For example, studies have investigated the role of blogs in migrants' adjustment to Canada (Nardon et al., 2015), using cell phone data to track population movement following disasters (Bengtsson et al., 2011), and how immigrants contribute to and leverage information on social media (Zijlstra and van Liempt, 2017). While these studies contribute to understanding the connections between ICT and migration, they do not provide a big picture of how and where these technologies can be helpful or what future improvements to technology have more potential for migration purposes. Existing studies are also from specific disciplinary (commonly non-technical) perspectives. While the researchers may be familiar with the technology, their analyses are usually from a user's point of view rather than a technology innovator, designer, or developer. This user-perspective limits their ability to envision a broader picture of how the technology can contribute to the migration process.

In this paper, we contribute to the field of technology and migration by proposing a framework for technology application (with existing tools), development (of new tools), and research (on novel features). We aim to support migration researchers and practitioners to understand the potential of emerging technologies use and development to address migration challenges. We take into account the technological affordances, the various users, the specific types of digital media to be used, and the migration-related purposes for which technology can be used. While we limit our discussion in this paper to ICT, we admit that other forms and areas of technology have significant connections to the migration process that are beyond the scope of this paper that aims at briefly present a conceptual model of technology use. We acknowledge that migration stakeholders follow different and sometimes conflicting agendas. Various political, economic, and social goals can be involved in the migration process and the use of ICT. For example, border crossing sees a wide range of technologies utilized by governments, migrants, and others involved with a range of goals, from preventing illegal migrants to smuggling them into a country. The purpose of our proposed framework is not to support any agenda but to offer a neutral understanding of ICT processes and how they link to different categories of migration tasks. We also recognize that in many cases, the limited use of ICT is not due to lack of knowledge of the potential but practical difficulties such as reaching out to migrants, implementation costs, and political restrictions. Discussing those topics is critical but beyond the scope of this paper.

In the following sections, we describe the proposed framework, examine some typical examples of technology use in migration, and discuss future possibilities. It should be noted that there are various forms of migration with their own specific issues. In this paper, we try to offer a technology use model that can, to some level, apply to all these cases. Also, the existing problems with the use of technology (many of them mentioned in the following sections) raise certain objections to more computer-based solutions and systems. While we understand these concerns, they are beyond the scope of this paper with its aim at offering a usage model.

## 2. Research Approach

### 2.1. Proposed Migration-Technology Framework

A conceptual model of technology use in migration is only possible to develop through an interdisciplinary lens that brings together not only technology and migration, but also business, communication, geography, and other social and health sciences. In an attempt to establish a manageable theoretical framework for our model, we organize the framework around two questions:

1. Who are the primary users of migration-related technologies? Users are the stakeholders in the migration process who perform activities that directly or indirectly affect migrants.
2. What are the common technical processes involved? Technical processes are the building blocks of any ICT system, regardless of its purpose.

Our framework aims at mapping tasks (the uses of technology for a migration-related purpose) to the users involved and technical processes performed. This will allow us to identify gaps and possibilities for new applications of existing technologies, customizing them for migration purposes, and research on new technological features for migration purposes.

### 2.2. Users of migration-related technology

We adopt a relational perspective on migration (Syed, 2008) which suggests that migrant outcomes are influenced by macro (national), meso (organizational), and micro (individual) levels. Such multi-level perspectives have been successfully proposed and used by other researchers (e.g., House, Rousseau, & Thomas-Hunt, 1995; Lee et al., 2020), and offer advantages over single-level models as they provide a hierarchy (Syed, 2008). In the context of technology and migration, we consider the macro level to encompass technology users involved in governance and policymaking and macro-level decisions that affect either migrants or the nature and structure of migration processes. At meso level, we include other organizational actors, such as community or not-for-for profit organizations directly supporting migrants as well as for-profit organization serving or employing migrants. The micro level includes both individuals and families going through the migration experience. The primary users of migration-related technologies by level are represented in Figure 1.

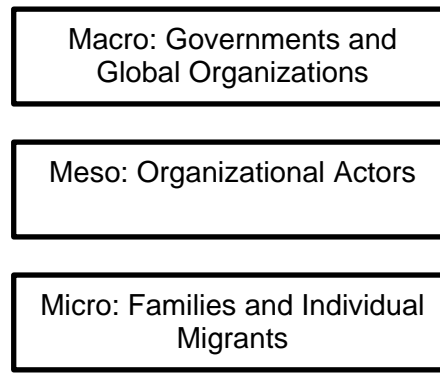

Figure 1. Three-level model of migration-related technology users

Users at each of these three levels engage in a series of tasks corresponding to different needs. While these needs may seem common at different levels, they are approached differently. For example, migrant employment is a concern at all three levels. While the micro-level users search for existing employment options to choose the most appropriate one, the meso-level users may focus on establishing and linking databases of needs (migrants looking for a job) and resources (available positions and employers). Macro-level users also deal with employment, but their efforts are focused more on reshaping the employment scene through policies, national priorities, and funding plans. Similarly, travel, housing, education, connections, and well-being are common needs that are dealt with at all levels. Satisfying these needs is the basis of migration-related tasks (activities) which can use existing, customized, or novel ICT tools. We refer to these three as application, customization, and research, respectively.

## 2.3. Technical Processes

To understand the role of technology in migration, we focus on the technical processes involved. All information technologies are basically data processing tools. The concept of data is commonly defined within the context of the following hierarchy (Stafford, 2009):

- Data is the raw "bits and bytes" that numerically represent a subject.
- Information is a specific data used for a specific purpose.
- Knowledge is structured information.
- Wisdom (or intelligence) is the ability to use knowledge to solve a problem or make a decision.

Researchers have defined essential properties of digital media such as programmability, interactivity, multi-dimensionality, encyclopedic nature, and distributed behavior (Murray, 1998). An approach to the design of truly digital solutions (for example, for migration tasks) is to effectively utilize these properties. For example, when designing an educational app for migrants, this approach suggests personalization to utilize the programmability (dynamic behavior), or embedded socialization to utilize multi-dimensionality. But it may not provide a systematic way of understanding what ICT can do.

In the field of ICT, the von Neumann architecture (1945) is a well-accepted representation of a computing system (Figure 2). Based on this representation, we can state that ICT tools engage in data collection (input), data presentation (output), data transformation (various operation on the data), and data storage.

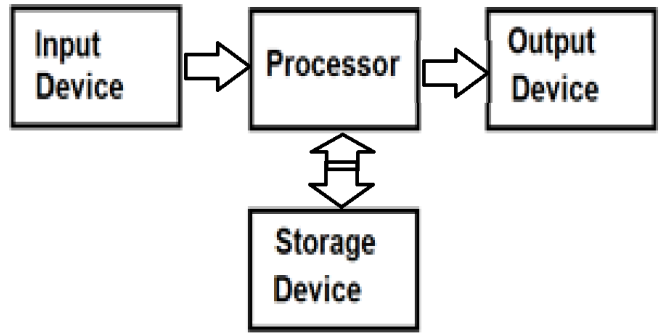

Figure 2. Generic Representation of von Neumann Architecture for Computing Devices

Inspired by von Neumann model, we suggest that ICT tools used for any migration-related activity perform one or a combination of the following technical processes on some data, potentially involving higher levels of data hierarchy:
1. Input or Collection: gathering existing data from different sources, including users, for example, through surveys or sensors
2. Creation: generating new data or transforming existing ones, for example creating maps
3. Analysis: processing collected data to find patterns, relationships, and other high-level information, for example identifying the correlation between events and migration patterns
4. Interpretation: making sense of data and understanding its meaning and impact, for example, studying what social media means to migrants
5. Output or Presentation: making data available to be used by a different audience, for example, through reports

Search and retrieval, storage, and transmission of data are among other common data processes that are used in many ICT systems, including migration applications. However, their use has been generic and not customized for migration. They have also not been studied in the literature in relation to migration. As such, we focus on the five processes listed above.

Table 1 shows some emerging technologies related to these processes.

| Technology | Related Processes |
|---|---|
| smart wearable sensors | data collections |
| smartphones | data collection and presentation |
| social media | data collection and presentation |
| artificial intelligence | data creation, analysis, and interpretation |
| augmented/virtual reality | data presentation |
| games and storytelling | data creation and other processes |
| secure systems such as BlockChain | data storage, creation, and transformation |
| crowdsourcing | data collection, creation, and transformation |

Table 1. Example of Emerging Technologies and their Processes

These technologies provide a series of affordances, i.e., properties that are compatible with and relevant to users' interaction and usage (Gaver, 1991). For example, virtual reality can offer strong senses of immersion and presence. Wearable sensors are among the essential tools for multimodal interaction by collecting data beyond the traditional mouse and keyboard. Social media are characteristically multi-dimensional as they can be used for communication, collaboration, and other purposes through text, images, audio, and video. Computer games and digital storytelling combine participatory and engaging experiences with entertainment and fun, while artificial intelligence allows recognition of patterns and trends in data to generate knowledge and facilitate decision-making and problem-solving.

### *2.4. Migration Technology Matrix*

As shown in Figure 3, we combine the users of migration-related information and technical processes involved in what we call the Migration-Technology Matrix (MTM). Each cell includes many migration-related tasks that share the same user group and data processes but vary in specific need and technology. For example, sentiment analysis and migration modeling are two tasks involving analysis processes done at macro level (same cell of the matrix), but serve different needs and use different ICT tools. Sentiment analysis is used to understand how people feel about migration issues, while migration modeling aims at understanding the patterns and trends of migration. Each of these tasks may be implemented using different technologies. For example, sentiment analysis can be done using surveys, analysis of news, or analysis of social media, as discussed later. A simple text table on a desktop computer may be used to present simple data, while an interactive three-dimensional (3D) visualization using an immersive head-mounted display (HMD) can be more helpful when dealing with large multi-dimensional data.

Some tasks are simple and fit into only one cell of the matrix (for example, the job search), while others cross levels or use more than one technical process (for example, a job search tool that allows customization of search results and presenting them in different forms). The proposed Migration-Technology Matrix (MTM) is the basis of a conceptual model that can map various affordances of emerging technologies to migration-related users and their needs and activities. We demonstrate MTM through a series of tasks that are identified in Figure 3. The tasks are identified for groups who use them more often.

| | Creation | Collection | Analysis | Presentation | Interpretation |
|---|---|---|---|---|---|
| **Macro** | | Sentiment Analysis | | Visualization | |
| | | Migration Modeling and Simulation | | | |
| **Meso** | Communication | | | | |
| | Education, Connections, Storytelling | | | Storytelling | |
| **Micro** | Information Exchange, Decision Support, Well-being | | | Well-being | |

Figure 3. MTM and Some Typical Migration-related Tasks

For each task, we review a series of cases in the literature that demonstrate typical technologies used for the technical processes involved. The cases are chosen to demonstrate the typical uses of emerging ICT tools. We will review the existing systems, identify strengths, weaknesses, and potential areas for improvement. The goal of our review is to illustrate how ICT can be applied for migration-related purposes to take full advantage of technological affordances, how customized tools can be developed, and what research goals can be defined to improve the role of technology in migration. We conclude the review with a series of suggested research questions and directions.

Our review is organized primarily using the three macro, meso, and micro levels for users. The cases we review almost always include more than one user level, but we associate them with levels based on the primary user. For example, if a web-based service is developed by the government to offer employment information, then it is associated with macro level because it is the government agency who has to deal with this tool first in order to design and develop it. On the other hand, if migrants form a network of blogs to share information, it is a micro-level tool even if organizations and government agencies later join and comment on blogs or add their own blog posts. The level boundaries are inherently fuzzy, and it is expected that some cases can be considered at more than one level.

3. **Technology at Macro-Level: Global and National**

Traditionally, global organizations such as UN/IOM and governments have used formal data sources such as census and registrations to collect migration data. In addition to their unique role as policymakers, they also provide services such as health, employment, and education through various agencies as well as serve as an information hub to migrants. Web-based services for data collection and information sharing are common among macro-level users. Emerging technologies have resulted in new data sources such as cellphone and GPS, and social media, in addition to new tools such as virtual reality and artificial intelligence. These new data sources and tools have changed the way many migration-related tasks are done. In the following sections, we start by reviewing the new source of data made possible by emerging technologies that affect all tasks that need data collection (primarily at macro but also meso and even micro levels). We then review some typical examples of new initiatives to use modern ICT tools in migration processes.

*3.1. Traditional and Emerging Data Sources*

The IOM's Global Migration Indicator (2018) recognizes the importance of migration data for decision-making, the problem of limited data, and various reasons that contribute to it. These reasons include quality and frequency of data collection, frequency of data sharing, irregular migration, and focus on only national levels. Supporting an extended set of data collection methods, and inspired by IOM, we can categorize data sources into two groups:
- Traditional data sources such as censuses, population registers, surveys, and border statistics.
- Innovative data sources such as social media, cell phone data, and online news content.

While census and population registers have been used to collect stock data (related to the number of people residing in a location), surveys can also provide flow data (related to movements), all with limitations due to collection frequency, inconsistencies in the data definitions, and poor quality of the data (Massey et al., 1999; Abel, 2015; Willekens et al., 2016; Willekens, 2019). Collecting, storing, and accessing this data is done through traditional ICT tools such as websites and desktop applications and various computational algorithms (Poulain et al., 2006; Abel, 2013; Kim and Cohen, 2010). Emerging ICT tools and techniques in recent years have provided alternative source of data that can remedy the limitations of the traditional ones, as described below.

**Cell Phone Data**

Mobile phones provide a new increasingly available source of data (Blumenstock, 2012). Their data often comes in the form of anonymized Call Detail Records (CDR) to researchers. CDR have demonstrated its usefulness in unearthing geographic mobility trends. Although, due to the anonymization process some information may be lost, CDR still contains useful spacio-temporal data that facilitates research on population movement. For instance, Bengtsson et al. (Bengtsson et al., 2011) leveraged CDR to track population movement following disasters. Researchers have shown the efficacy of CDR in capturing the behaviors of undocumented immigrants, temporary workers, circular migrants, among others (Neubauer et al., 2015). Leveraging CDR in some locations in Cote d'Ivoire, where there is a lack of granular data on population mobility, Lu and colleagues (Lu et al., 2013) found that they could estimate a potential predictability in user mobility as high as 88% by applying a Markov Chain-based estimation algorithm.

In addition to CDR, certain mobile phone features other than mobile network itself, such as Wi-Fi, GPS, Bluetooth can provide useful tracking information for measuring population movement. For instance, applications like Google Latitude allows users to track their movement and behavior by integrating data from Wi-Fi, GPS, and GSM localizations. Delafontaine et al. (2012) leveraged Bluetooth receivers throughout a convention space, which recorded the unique footprints of nearby Bluetooth devices as they passed within the range of their signals. Privacy is a major concern when using such mobile features (Musolesi, 2014; Shilton, 2009).

Despite various advantages of CDR in tracking population movement, it still has many limitations. The most significant being the unreliability of cell tower information (Csaji et al., 2013; Bayir et al., 2010). Another issue with CDR is the inconsistencies of cell phone usage in rural to urban landscape, which may not properly capture the entire population of a large region (Blumenstock and Fratamico, 2013). Also, anonymized CDR does not provide much demographic information, which makes it difficult to understand users' residency status, which in turn poses difficulty in drawing any conclusions from CDR on international migration.

**Online News and Articles Data**

Large number of online articles from past time periods facilitates quantitative analysis, especially, text analytics, which allows identifying how migrants or the topic of immigrations are presented in the news media. For instance, Schmitz, N. J. (2019) examined articles of one Dutch newspaper related to immigration published between 2013 and 2019. Their findings suggest that the average sentiment of these articles was in general negative. Another group of researchers performed quantitative discourse analysis on 15 years (1999-2013) of Dutch newspaper content on unauthorized migrants in order to understand if media content fuels the *crimmigration* process (Brouwer, J., van der Woude, M., & van der Leun, J. 2017).

Hadiji et al. (Hadiji et al., 2013) studied migrations of academic researchers by leveraging bibliographic information from the DBLP computer science bibliography. Journal publications records information about each author and their associated institution. This allowed observing the movement of researchers from one institution to another, and eventually to study the aggregate mobility of academic researchers.

**Social Media Data**

Increased usage of social media platforms such as social networking programs and blogs provide a new source of information. Despite the inaccuracy and inconsistency (Charmarkeh, 2013), the vast amount of data can still provide significant insight.

Zijlstra and van Liempt (2017) demonstrated that immigrants contribute to and leverage information on social media. For instance, social media can impact decisions of immigrants about the routes to take, the choice of transportation (e.g. plane, boat, or on foot). Dekker et al.'s (2018) performed analysis on social media use by asylum migrants prior to and during migration based on in-depth interviews with 54 Syrian asylum migrants who recently obtained refugee status in the Netherlands. Their findings reveal that the majority of Syrian asylum migrants had access to social media information before and during migration which means social media can be used to collect data from migrants or to provide information to them.

As immigrants stay connected with friends and family through mobile phone and social media programs during and after their move to the destination country, data generated in such platforms often include information on their travels and country of origin and destination (Diminescu, 2008). Geo-localized social media data can be used to identify routes taken by migrants, to analyze migration flows in near real time. Data generated on social media may even indicate why people choose to migrate from one place to another, the migrants' living conditions in destination countries vs. countries of origin, how migrants integrate with the communities in the destination countries, and the public forms attitudes towards the immigrations.

Alexander et al (2020) proposed a framework to combine social media data with traditional survey data to produce timely 'nowcasts' of migrant stocks by state in the United States. State et al. (2014) leveraged a million geo-tagged career histories provided by LinkedIn in order to estimate the number of professional migrations in the United States.

Herdağdelen et al. (2016) leveraged migrants' Facebook links to evaluate their degree of integration in the United States by nationality and if they lived in areas with concentrated population from specific countries. Their result shows that migrant groups with cultural proximity had a similar degree of integration. Dubois et al. (2018) compared cultural assimilation among migrants in Germany between migrants from Arabic-speaking countries and migrants from other European countries by looking at "likes" migrants shared online for a list of interests on Facebook pages. Their results demonstrate that Arabic-speaking migrants had a lower assimilation score than other European groups. The language used in social media posts can also help predict cultural assimilation. For instance, Lamanna et al. (2018) analyzed geo-tagged tweets from 53 of the most populated urban areas of the world between 2010 and 2015. The comparison between the language in which the user posted and the language prevailing in the area allowed constructing a degree of integration of migrants in their surrounding societies.

A common limitation in most studies that used social media data for estimating migration lie in the sample data used for analysis. While many such studies demonstrated results to closely resemble official data for the same measurement, how data was selected still may result in bias in the study. This is especially true for social media data, where users may not always share the actual and current information about their whereabouts. Moreover, social media users of a particular platform may not necessarily represent the entire population. For instance, in State et al.'s (2014) study on using LinkedIn to estimate highly skilled migration to the US, the researchers did not have a way to know the citizenship status of individuals in the sample. As a result, their dataset may not properly differentiate between the return migration of US expatriates and in-migration of foreign persons. In addition, as LinkedIn users are not a representative sample of the entire population of highly skilled migrants, the estimates derived from this data may be biased.

### *3.2. Sentiment Analysis*

Several multilingual dictionaries or lexicons, such as Sentiwordnet (Esuli & Sebastiani, 2006) or Multiwordnet (Pianta et al., 2002) exist that allow performing sentiment analysis on the textual content of social media. For instance, Mohammad et al. (2013) have developed a specialized lexicon that supports emotion extraction. Chung and Zeng (2016) developed a framework for such analysis that consists of various steps of data collection, cleaning, filtering, and analysis (Figure 19). The framework uses Social media data as input, a specialized lexicon for emotion extraction and sentiment analysis, an algorithm to compute emotion indices, and program modules for performing network analysis and visualization. Based on this framework, they developed a system called "iMood" that performs sentiment and network analyses of U.S. immigration and border security. They analyzed 909,035 tweets posted by over 300,000 users during three phases between May and November 2013 using iMood, and investigated the sentiment, emotion, and network characteristics found in those tweets. Their analysis reveals significant differences in emotion and sentiment among all three phases (Figure 20). They also identified influential users based on their influence scores derived through network analysis across the three phases.

By analyzing 2 million tweets containing the word "refugees" (in 6 different languages, sent between 2015 and 2017), Gualda and Rebollo (2016) highlighted that discourses were emotional and polarized. Some tweets were pro-refugee messages which were critical towards governments, while some other tweets were also openly racist towards the migrants. Since the increase in immigration during 2015 and 2016, social media have hosted a fierce political debate on immigration issues. The total number of tweets in Italia on this subject has increased sharply. The Italian institute of statistics developed a tool to perform regular sentiment analysis on Twitter messages shared by Italians (Righi et al., 2018). It measured the daily mood of users and the effect of broadcast headlines regarding migrations.

Textual analysis has its limitation when it comes to identifying sentiment. The process for selecting sample words or terms must be robust as a non-random selection of the terms can significantly affect the results. The messages shared on social media on one topic can also vary by demographical differences, such as sex, age, and nationality (UN Global Pulse and UNHCR Innovation Service, 2017). By using machine learning algorithms, one group of researchers showed that people may not express their actual sentiment about immigration on Twitter (Bartlett and Norrie, 2015). Rather, they may often react to events or complain about the content of other messages on the media. A more robust filtering mechanism for sample collection may help reduce the impact of such inconsistencies in data and subsequently provider more accurate results.

Using large quantitative data also has the drawbacks of relying on only one source of information and missing on subtleties that can only be noticed from qualitative data. Using multiple data sources, and identifying the type of information that they are suitable for are open research objectives in sentiment analysis. Using AI methods for combining and analyzing qualitative and quantitative data from multiple sources is a promising research direction for sentiment analysis.

### 3.3. Migration Modeling and Analysis

Traditional statistical methods have long been used to model migration data (Haan and McDonald, 2018). In addition to Natural Language Processing (NLP) (Indurkhya & Damerau, 2010)for sentiment analysis, other artificial intelligence and data science methods, especially deep learning (Goodfellow et al., 2016), can also be used to model various aspects of migration for decision-making purposes. Social media plays a key role in understanding the migration patterns of individuals and demographics of a place (Dekker and Engberson, 2012). Machine learning analysis of Twitter users data their social network structure is used to predict ethnicity and other attributes (Pennacchiotti & Popescu, 2011; Rao, 2010). Saravanan (2017) used the tweets of users belonging to a particular ethnicity/race as a training set for preparing a deep learning neural network model (Shen, 204) and test the user samples for categorizing them into different ethnicity.

Basile et al. (2019) reviewed a series of intelligent methods including deep learning to identify hate speech against migrants and women in tweeter. Wang et al. (2018) used deep learning for urban resident recognition. Overall, the few efforts to use AI and machine/deep learning for migration data analysis have been promising and there seems to be a strong potential for further research in this area. These can be to model various migration trends for prediction purposes or recognize and classify patterns to support evidence-based decision making.

### 3.4. Visualization

In addition to analyzing data to find patterns and trends, new data science methods help with visualizing information. New software tools such D3 (https://d3js.org), and IPython (https://ipython.org), combined with machine learning programming libraries such as Scikit-Learn (https://scikit-learn.org) and TensorFlow (https://www.tensorflow.org) provide researchers and others working with large databases the ability to create interactive visualizations that are more effective in giving insight into the collected data especially (Weiskopf, 2007; Tominski and Schumann, 2020). Such interactive visualization is particularly helpful when working with large and high-dimensional data (Godfrey et al., 2016; Duque et al., 2019).

Newer trends in visualization include the use of Augmented Reality (AR) and Virtual reality (VR) to provide immersive visualization (Su et al., 2019; Tadeja et al., 2020; Quach and Jenny, 2020; Klein, 2019; El Beheiry et al., 2019). In a review of projects using different visualization systems, Rozmi et al. (2019) concluded that non-immersive visualization tools (both 2D and 3D) have "drawbacks such as unreliability, high cost, and lack of immenseness." They found immersive VR visualization particularly helpful for cases that "require instant feedback" as the user can easily navigate through the data and interact with it. Rozmi et al. also emphasized the importance of combining immersive visualization with advanced machine learning algorithms and procedural generation of 3D environments based on the data and user interactions.

Very few efforts has been made to use the recent advances in visualization for migration studies. Wihbeyet al. (2020) move away from traditional visualization forms and use the visual metaphor of tree to represent the US population diversity. Recognizing how the collection of administrative data and its rendering are political acts, they "Data visualization that is rigorously rooted from a computational and statistical perspective, while at the same time innovative in generating ideational and compositional meanings, can help transcend limitations of administrative data and produce new discourses about diversity and its importance in society." Risam (2019) also emphasizes on the political nature of data and visualization, and compares two approaches: "(1) taking existing data sets and creating visualizations without collaboration with the entities that collected the data or (2) with migrants themselves and cocreating visualizations with migrants through participatory projects." Seeing visualization as a narrative, Risam emphasizes the importance of the participatory design (with migrants themselves) for visualizations to avoid the "migrant-as-problem" perspective. Recent approaches to narrative-based Human-Computer Interaction (Wright & McCarthy, 2008; Tanenbaum, 2014) and the participatory abilities of digital media to create more inclusive visualization, allowing migrants to tell their stories within the context of data visualization, have not been studied properly in the field migration and offer a strong research opportunity. Initiatives to offer migrants a platform to tell their stories have been generally for community connection, empathy, and empowering/healing purposes as we see in the next section. While valuable, these efforts have not been integrated into any decision-making process at macro level.

### *3.5. Communication with Migrants*

Governments and migration-related organizations have long used traditional ICT such as websites and email lists for communication with migrants. More recently, governments and NGOs have moved towards using social media a communication tool (Beyer, 2017; Underhill et al., 2019; Dekker et al., 2018; Alencar, 2018). Widespread access to mobile devices makes the use of social media more feasible but the lack of access and trust, an also fear of government control (Miller, 2018), are among issues preventing the effectiveness of such efforts as migrants rely more on their own trusted networks (Pannocchia et al. 2020). Using specially designed apps and sharing and collecting data through them are other potentials that can be considered but are less used (Gros & Maniega, 2018; Bustamante Duarte et al., 2018; Bock et al., 2020). We will review some of these apps, developed by different organizations, in the next section. Privacy and trust concerns are again issues that need to be addressed before such methods can be widely used. Co-design and participation of migrant in the design of these apps have also been suggested (Gros & Maniega, 2018). Wearable technologies such as smart bracelets, RFID, and even regular bracelets with QR code are other candidates for communication (Wister et al., 2015; Lamber et al., 2018) that are not seriously pursued in the field of migration yet.

**4. Technology at Meso-Level: Organizations and Communities**

Various NGOs and community organizations have been working with migrants on issues such as employment, housing, language, childcare, and social connections. The developed technologies include features such as supporting newcomer navigation, helping them find work, and offering access to community housing and services (Benton and Glennie, 2016). Similar to governments and global organizations, these groups have used traditional Internet technologies (mostly web-based) for communicating with migrants.

To leverage the increasing use of mobile devices among migrants, a few organizations have also developed mobile apps (Buono et al., 2019; Costa, 2018). For example, RefAid (http://www.refaid.com) mobile app shows required services to migrants on a map with a simple interface. Trusted aid organizations can manage and update their services using a web-based content management system. Other apps such as MigAdvisor and ImMigRant also provide immigration and emergency advice through simple mobile interfaces. Relying on community information through crowdsourcing (Taeihagh, 2017), connecting to existing social networks, and using distributed trust systems such as blockchain (Crosby et al., 2016) are among future possibilities when using mobile apps to provide trustworthy information to migrants. In addition to general communication, establishing connections, empowering and healing, and education are among the areas where emerging technologies are (or can be) proposed at meso level.

### *4.1. Establishing Connections*

Social capital is defined as "the acquaintances, friends and family members that make up our safety net," (Woolcock, 2001; Garang, 2012) and those who have a higher social capital are better able to confront poverty and vulnerability, resolve disputes, adapt to cultural norms, and take advantages of new opportunities. Studies show that many migrants, particularly women, suffer from a low social capital (Houle & Schellenberg, 2010; Hudon, 2015). Considering that the migrant's support-seeking strategies rely more on informal networks such than formal ones provided by governments and NGOs (Stewart et al., 2008), migrants have a strong need to increase their social capital. Many meso-level groups try to offer migrants with the opportunity to connect to others in their new home community.

**Crowdsourced Networking**

In addition to more traditional methods such as mentoring, networking events, and publications, technology-enabled methods of connection have been used by some organizations based on crowdsourcing and "sharing economy" approach (Cukier & Jackson, 2017; Cukier, 2016). One project in particular, the Ryerson University Lifeline Syria Challenge (RULSC), used crowdsourced donations and organized volunteers to help the families resettle, as well as finding housing and jobs through the platform Magnet (https://magnet.today/). The project sponsored 75 families (300 newcomers). Importantly the donations ranged from $1,000,000 to $10. Cukier states that the project was able to bring out people other than the "usual suspects" that tend to be involved with private sponsorship. Crowdsourcing (Taeihagh, 2017) is a fairly recent development enabled by ICT. It is made possible through interactive personalized mobiles apps and web services such as Foursquare City Guide, a crowdsourced travel and navigation app. Crowdsourcing provides many potentials (Reis et al., 2016; Gebert, 2016) such as job finding, problem-solving, and social connection, which are not well investigated and utilized in the migration field.

**Virtual Reality for Empathy**

Establishing connections can be facilitated through interactive tools, but it cannot succeed without an emotional connection between two parties. In addition to its potential for visualization (Rozmi et al., 2019) and education (Scavarelli et al., 2020), it has been suggested that virtual reality can be a tool for increased empathy towards certain groups by creating an immersive experience (Herrera et al., 2018; Bertrand et al., 2018). Hunger in LA (https://emblematicgroup.com/experiences/hunger-in-la/) and Carne-y-arena (Flesh and Sand,

https://www.lacma.org/art/exhibition/alejandro-g-inarritu-carne-y-arena-virtually-present-physically-invisible) are two examples of well-received VR experiences designed to let audience emotionally connect to the subject. The former simulates the experience of helping a diabetic man who collapses at a food bank, and the latter is about illegal immigrant crossing the US/Mexico border.

While the long-term effect of "VR empathy machines" is questioned, various recent efforts have successfully investigated the use of VR with regard to migrants (Bystrom & Mosse, 2020; Keating, 2017; Kros et al., 2016; Alberghini, 2020; Auchter, 2020. Camilleri et al., 2018; Camilleri et al., 2016). The results have indicated that VR's immersive quality "can enhance the content by eliciting both emotional and cognitive empathy and help develop perspective-taking capacities" (Alberghini, 2020). They also show that, at least in the short term, "the empathic reaction stimulated some forms of altruism expressed in the commitment to offer a personal contribution". These VR experiences can be combined with gaming elements, Augmented Reality (AR), and other immersive, intelligent, and interactive to offer more potentials. Such use of VR has strong potential to establish connection to migrants and is a subject of future research. Longer-term studies, increasing visceral engagement, and reflective abilities are among particular areas that require such further research.

## *4.2. Empowering Migrants through Storytelling*

In previous sections, we discussed the need for migrants' participation in the process of migration data visualization. We also saw how the immersive experience of migrants' narratives can help increase empathy and social connections. Storytelling can also have an empowering effect for migrants themselves. Migrants public narratives can challenge and revise the master narratives (McLean & Syed, 2015) and create a more inclusive society (Moffitt et al., 2020). Storytelling can also be a personal sense-making process (Frank 2013; Degat et al., 2016). It is argued that digital storytelling, through the use of multimedia and interactive methods, provides a more empowering way of sharing experiences, especially for marginalized people (Gubrium, 2009; Saridaki, 2018, DeVecchi et al., 2016).

Online and digital storytelling forums for migrants have been set up by both governments and various organizations. They use a variety of formats from text such as Passages Canada (http://passagestocanada.com) to video such as Digital Storytelling by Canadian Museum of Immigration (https://pier21.ca/digital-storytelling) and interactive workshops (https://www.afsc.org/resource/how-manual-digital-storytelling-immigrants-and-refugees). Wieland et al. (2017) report on the success of digital storytelling interventions among immigrant populations in primary care settings which resulted in self-rated health improvements. McDonough and Colucci (2019) also report on the success of an immersive digital storytelling workshop for 10 immigrants with mental health issues and state that "Individuals experienced the project as personally empowering. The stories have been used to facilitate community and service-based conversations about diversity, mental health and recovery."

Johnson and Kendrick (2017) present a digital storytelling project conducted in a school district's transition program, where adolescent migrants used a multimodal approach to shared aspects of their identities and social worlds. The study finds the effectiveness of nonlinguistic modes for exploring and making visible complex and abstract facets of their life and identity. This is in line with new technological trends towards multimodal Human-Computer Interaction (Turk, 2014) and suggests a potential, less explored, venue to increase the effectiveness of storytelling. Advances in computer game design and interactive storytelling approaches (Glassner, 2017) can also be utilized in this regard.

### *4.3. Education and Adjustment*

Education and training have long been a significant part of efforts by governments and migration-related organizations (Hannah, 2008; Wrigley, 2007). Language skills are among the most common educational goals (Wrigley, 2077) while employment-related and social skills are also pursued (Nakhaie, 2018; Donitsa-Schmidt, 2017). These programs use traditional ICT in the form of websites, webinars, online education, and similar tools and are primarily for adults.

A variety of support services and programs exist for newcomer children (Olsen, 1989; SPCO, 2010). They range from counselling (Kopala et al., 1994), various community-based activities and support groups (Aronowitz, 1984), to storytelling (Rousseau and Heusch, 2000) and creative workshops (Rousseau et al., 2005; Miller and Billings, 1994). The role of educational support to facilitate children adjustment to foreign environments is well established (McBrien, 2005). There is also research suggesting the value of expressive, engaging, and entertaining activities for education of immigrant and refugee children (Miller and Billings, 1994; Rousseau et al., 2005). However, researchers have rarely explored the use of digital media and interactive entertainment in facilitating this adjustment. A strong stream of research suggests that digital media and interactive technologies have a positive impact on educational outcomes (Connolly et al., 2012; Hwang and Wu, 2012). Storytelling has been known as an effective educational method (Hamilton & Weiss, 2005) and many educators have utilized digital and interactive storytelling in order to provide information in a more engaging way (Ohler, 2013; Yang and Wu, 2012). Computer games have been used to assist children with special educational needs (Durkin et al., 2015), cultural and behavioural learning (Kirkland & Wiliamson, 2010), develop cognitive skills (Blumberg and Fisch, 2013), and provide learning experiences in many different fields and for various age groups (see Connolly et al., 2012, and Hwang and Wu, 2012, for some examples).

Arya et al. (2014) present the results of using 3D games and virtual environments for language training of international students. Bani-Taha et al., (2019) have reported one of the few works on using digital games to educate migrant children. This work, in line with our discussion on storytelling and immersive experiences, demonstrate the potential of game-based learning and interactive storytelling as a method of training and education for migrants, especially children. As in the case of visualization, the role of participatory design to include the migrants is critical but not investigated enough.

## 5. Technology at Micro-Level: Families and Individuals

Thanks to the availability of the Internet and mobile services, a growing number of migrants are using ICT tools from early contemplation to long-term integration stages of their migration process (Pannocchia et al., 2020). Technology supports them by promoting language learning and connecting to various opportunities (Dekker and Engbersen, 2014; McGregor and Siegel, 2013) while retaining their cultural identity and maintaining their connections with their home countries (Almohamed and Vyas, 2016). Pannocchia et al. (2020) performed a literature review and found that the key areas of technology use by migrants included gaining information (for navigation and decision-making), establishing and maintaining connections. Within their dataset, smart phones were the top hardware used by migrants while computers and standalone GPS devices were used much less. Facebook, Twitter, and WhatsApp were the top software platforms while Skype and Instagram were also used. Other research (Nardon et al., 2015; ) also shows that blogs, discussion forums, and other social and navigation apps are also heavily used, and sense-making and well-being are among salient purposes of technology by migrants.

### 5.1. Information Exchange and Decision Support

Nardon et al., (2015) identifies technology as supportive to individuals adjusting to a new country. They found that blogging (combination of bloggers, readers, technology, and content) provides online support resources for adjustment. Such online support can help migrants "deal with their experiences of uncertainty, ambiguity and anxiety and include information, interpretation schemas, and comfort." Similarly, discussion forums (Gulanowski, 2018) and communication apps such as WhatsApp and Telegram (Pannocchia, 2020) can play significant roles for migrants to create social support.

These tools are generally not customized for migrants use. For example, the ability to link communication to navigation, identify points of interest on a map, and allowing comments from other migrants to help access these points can be very useful. While some organizations are developing specialized apps (for example, RefAid discussed earlier), there are also practices that can be used to integrate existing tools. Informing migrants of bets practices to use technology can be an area of future work.

While many ICT tools are being used for collecting and exchanging information, fewer exist that help with decision-making through interpretation of information or making sense of experiences. Migrants use communication tools (including blogs and storytelling) to achieve these goals but there is potential in intelligent and interactive technologies to help with reflection, data interpretation, and similar activities.

### 5.2. Well-being

Similar to more advanced cognitive tasks such as sensemaking and interpretation that are not properly covered by exiting tools, there is a lack of studied computer-mediated methods for migrants to help with their general well-being. Yao et al. (2019) discuss the (potentially) positive effect of social media on migrants' mental health when it is used to provide counselling services. While there are evidences that games, virtual reality experiences, and other computer-mediated activities can have positive effect on overall well-being (Carl et al., 2019; Fish et al., 2018; Gamage and Ennis, 2018; Sano and Picard, 2013), there are no studies that investigate using these methods for or by migrants.

6. **Discussion**

The Migration-Technology Marix (MTM) allows us to see technology-migration connections in terms of three levels of migration-related stakeholders and five main information processing types. As demonstrated in previous section, despite the increasing use of ICT by different migration stakeholders, the emerging technologies and their affordances have not been effectively used in this area. Table 2 shows a mapping of discussed technological solutions within the MTM framework. In this table, each cell of the matrix includes the ICT tools and methods that are well-utilized and also those that have significant opportunities for further application, customization, and research, as defined at the end of Section 2.1.

|  | **Creation** | **Collection** | **Analysis** | **Presentation** | **Interpretation** |
|---|---|---|---|---|---|
| **Macro** | **Well-utilized ICT tools and methods** | | | | |
| | Web-based, Social media, Some apps | Traditional and emerging sources | Statistical methods | Traditional presentations and web-based distribution | |
| | **Opportunities for further application, customization, and research** | | | | |
| | Specialized apps and social media presence, crowdsourced data creation | Wearables and other mobile data | More AI | Interactive, immersive, and intelligent visualization, integration with mobile and social apps | Limited (almost no) interpretation tools. Most interpretation of data is done "manually" by experts or non-expert users |
| **Meso** | **Well-utilized ICT tools and methods** | | | | |
| | Web-based, Social media, Some apps | Some surveys | Some statistical methods | Traditional presentations and web-based distribution | |
| | **Opportunities for further application, customization, and research** | | | | |
| | Specialized apps and social media presence, crowdsourced data creation | There is very limited data collection at Meso level | There is very limited data analysis at Meso level | Interactive, immersive, and intelligent visualization, integration with mobile and social apps | Same as Macro |
| **Micro** | **Well-utilized ICT tools and methods** | | | | |
| | Extensive use of emerging methods | Traditional and emerging sources | | | |
| | **Opportunities for further application, customization, and research** | | | | |
| | Storytelling Co-design | Better practices and integration of tools | Limited (almost no) data analysis tools | Limited (almost no) presentation tools | Same as Macro |

Table 2. MTM Mapping of Existing Technology-based Migration Tasks

The application opportunities address the effective use of existing generic ICT tools. For example, considering the extensive use of mobile devices by migrants, an effort should be made to provide content through new and existing mobile apps, and integrated with existing mobile services such as maps. Similarly, with the growth of new tools, crowdsourcing has been effectively used in many areas and could become integrated with migration services and practices, particularly for collecting and creating information. Wearable technologies have also been shown to be helpful in health and education tasks (for example, for monitoring, reminders, and notifications). However, their use in migration-related projects has been very limited and can be investigated more, not only in terms of using existing tools but also for customization and research.

The customization opportunities involve the development of ICT tools based on the existing features and capabilities but customized for migration tasks. For example, as mentioned above, wearable technologies have been customized for tasks such as health and well-being (activity trackers are a good example). They can be similarly used in tools customized for migration tasks such as data collection or communication. Digital storytelling and game-based learning have the potential for the education and well-being of migrants. Educational games and storytelling experiences need to be developed within the migration context to make use of these technologies. Another customization opportunity involves various data science and AI methods, particularly deep learning and machine learning. New tools can be developed based on these methods optimized for different types of migration data. Similarly, customized interactive, immersive, and intelligent visualization tools can help understand the large amount of migration data better. Search and retrieval, storage, and transmission of migration data currently follow the common methods in ICT with no significant customization. These areas are also open to customized development and research in the future.

Research on novel features and methods can improve various technologies and make them more suitable for migration tasks. For example, more studies are needed to investigate the effect of educational games and stories, also to find proper ways of including migrants in the design and development of such content. Such participation of migrants should also be investigated in creating, analyzing, and presenting different types of data for analysis and visualization. On the other hand, quantitative "big data" methods do not necessarily provide the proper insight into migration issues. More research is needed to identify how to define and use "good data" which can be qualitative involving direct participation of migrants.

There are topics that have received very little research attention within the migration context. The issues of trust (for example, who is a trustworthy source for migrants) and privacy (how to protect migrant's personal information while collecting useful data) still include many unknowns, and the procedures involved in data collection and communication have to be studied with regards to these issues. There is almost no ICT support for data interpretation or migrants' well-being. Intelligent methods have been used (Puiu et al., 2016; Roda et al., 2016) to help make sense of large amounts of data, but (1) they have not been investigated for migration data, and (2) they have not been developed for individual migrants who have to struggle with the information they receive. Last but not least, personalization and contextualization can be helpful in such cases as used in many intelligent systems recommending products and services to users (Aggarwal, 2016). A common trend in recommender and search systems these days is to provide recommendations (or search results) that are customized to the needs and interests of specific user or conditions of a specific context. Providing migration-related services to users at any of the here levels and based on the notions of personalization and contextualization is an open area of research.

## 7. Conclusion

We started this paper by introducing the Migration-Technology Matrix (MTM) as a framework to map out possible links between migration-related user groups (macro, meso, and micro) and technical processes (data collection, creation, analysis, visualization, and interpretation). Knowing common tasks the user groups perform and common technologies and affordances associated with the processes, we were able to (1) review a series of typical cases that used different ICT tools to perform those common tasks, and (2) identify the gaps in terms of application, customization, and research opportunities.

Our review was not exhaustive and only included examples of some typical tasks and technologies. For practitioners and researchers involved in those tasks, our suggestions can be helpful as we define potential future directions for their research and development activities. For those who are involved in other tasks, our framework can provide a way to identify such research and development opportunities themselves to effectively use the affordances of emerging technologies.

**References**

Aggarwal, C. (2016). Recommender Systems: The Textbook. Springer.

Alberghini, D. (2020). Improving empathy: is virtual reality an effective approach to educating about refugees?.

Alencar, A. (2018). Refugee integration and social media: A local and experiential perspective. Information, Communication & Society, 21(11), 1588-1603.

Almohamed, A., & Vyas, D. (2016). Vulnerability of Displacement: Challenges for Integrating Refugees and Asylum Seekers in Host Communities. Proceedings of the 28th Australian Conference on Computer-Human Interaction (OzCHI'16), 125–134

Aronowitz, M. (1984). The social and emotional adjustment of immigrant children: A review of the literature. International Migration Review, 237-257

Arya, A., Hartwick, P., Graham, S., & Nowlan, N. (2012). Collaborating through space and time in educational virtual environments: 3 case studies. Journal of Interactive Technology and Pedagogy, 2.

Auchter, J. (2020). Augmented Reality: Memorializing Deaths of Migrants Along the US-Mexico Border. In Public Memory in the Context of Transnational Migration and Displacement (pp. 267-291). Palgrave Macmillan, Cham.

Bani-Taha, O., Malek El-Kouzi, Ali Arya, Fraser Taylor, (2019). Educational Computer Games for Social Adjustment of Newcomer Children. In Proceedings of ICERI 2019

Basile, V., Bosco, C., Fersini, E., Debora, N., Patti, V., Pardo, F. M. R., ... & Sanguinetti, M. (2019). Semeval-2019 task 5: Multilingual detection of hate speech against immigrants and women in twitter. In 13th International Workshop on Semantic Evaluation (pp. 54-63). Association for Computational Linguistics.

Bengtsson, L., Lu, X., Thorson, A., Garfield, R., & Von Schreeb, J. (2011). Improved response to disasters and outbreaks by tracking population movements with mobile phone network data: a post-earthquake geospatial study in Haiti. PLoS Med, 8(8), e1001083.Benton, M., & Glennie, A. (2016). Digital Humanitarianism: How tech entrepreneurs are supporting refugee integration.

Bertrand, P., Guegan, J., Robieux, L., McCall, C. A., & Zenasni, F. (2018). Learning empathy through virtual reality: multiple strategies for training empathy-related abilities using body ownership illusions in embodied virtual reality. Frontiers in Robotics and AI, 5, 26.

Beyer, A., Brekke, J. P., & Thorbjørnsrud, K. (2017). Communicating borders: Informing migrants and potential asylum seekers through social media. Rapport-Institutt for samfunnsforskning.

Bock, J. G., Haque, Z., & McMahon, K. A. (2020). Displaced and dismayed: how ICTs are helping refugees and migrants, and how we can do better. Information Technology for Development, 1-22.

Buono, P., Cassano, F., Piccinno, A., Rossano, V., Roselli, T., & Berni, F. (2019). Multimedia technologies to support delivery of health services to migrants by enhancing their inclusion. In 2019 23rd International Conference Information Visualisation (IV) (pp. 370-375). IEEE.

Bustamante Duarte, A. M., Degbelo, A., & Kray, C. (2018). Exploring forced migrants (re) settlement & the role of digital services. In Proceedings of 16th European Conference on Computer-Supported Cooperative Work-Exploratory Papers. European Society for Socially Embedded Technologies (EUSSET).

Bystrom, K., & Mosse, R. (2020). Inside the "Empathy Machine": Virtual Reality and Humanitarian Campaigning. In Writing Beyond the State (pp. 87-110). Palgrave Macmillan, Cham.

Caidi, N., Allard, D., & Dechief, D. (2008). Information practices of immigrants to Canada: A review of the literature. Unpublished Report to Citizenship and Immigration Canada.

Caidi, N., Allard, D., & Quirke, L. (2010). Information practices of immigrants. Annual review of information science and technology, 44(1), 491-531.

Camilleri, J., Scicluna, D., Briffa, V., Camiller, V., Dingli, A., & Montebello, M. (2016). If you aren't experiencing it, you're not understanding it! Using virtual reality to understand migrants.

Camilleri, V., Dingli, A., & Montebello, M. (2018). Empowering Citizens Through Virtual and Alternate Reality. In Virtual and Augmented Reality: Concepts, Methodologies, Tools, and Applications (pp. 720-732). IGI Global.

Carl, E., Stein, A. T., Levihn-Coon, A., Pogue, J. R., Rothbaum, B., Emmelkamp, P., Asmundson, G. J. G., Carlbring, P., & Powers, M. B. (2019). Virtual reality exposure therapy for anxiety and related disorders: A meta-analysis of randomized controlled trials. Journal of Anxiety Disorders, 61, 27–36.

Connolly, T. M., Boyle, E. A., MacArthur, E., Hainey, T., & Boyle, J. M. (2012). A systematic literature review of empirical evidence on computer games and serious games. Computers & Education, 59(2), 661-686.

Costa, S. B. (2018). The ICT Business: smartphones and the Integration of refugees in Europe (Doctoral dissertation).

Crosby, M., Pattanayak, P., Verma, S., & Kalyanaraman, V. (2016). Blockchain technology: Beyond bitcoin. Applied Innovation, 2(6-10), 71 Connolly, T. M., Boyle, E. A., MacArthur, E., Hainey, T., & Boyle, J. M. (2012). A systematic literature review of empirical evidence on computer games and serious games. Computers & Education, 59(2), 661-686.

Cukier, W. and Jackson, S. (2017). Welcoming Syrian Refugees to Canada: technology-enabled social innovation. Humanitarian Technology Conference (IHTC), 2017 IEEE Canada International, (2017), 32–36

Cukier, W. Social Innovation in Immigrant Employment. National Metropolis Conference, Toronto, (2016).

Culotta, A., Ravi, N. K., & Cutler, J. (2016). Predicting Twitter user demographics using distant supervision from website traffic data. Journal of Artificial Intelligence Research, 55, 389-408.

De Haas, H., Castles, S., & Miller, M. J. (2019). The age of migration: International population movements in the modern world. Bloomsbury Publishing.

De Vecchi, N., Kenny, A., Dickson-Swift, V., & Kidd, S. (2016). How digital storytelling is used in mental health: A scoping review. International Journal of Mental Health Nursing, 25(3), 183-193.

Degât, S., Negus, S., Pyzocha, A., & Rommerskirchen, S. (2016). DIFFERENT TAKES ON THE ARRIVAL: EMPOWERING IMMIGRANT STUDENTS THROUGH STORYTELLING.

Dekker, R., & Engbersen, G. (2014). How social media transform migrant networks and facilitate migration. Global Networks, 14, 401–418

Dekker, R., Engbersen, G., Klaver, J., & Vonk, H. (2018). Smart refugees: How Syrian asylum migrants use social media information in migration decision-making. Social Media+ Society, 4(1), 2056305118764439.

Dominguez, D., Soria, P., González, M., Rodríguez, F. B., & Sánchez, Á. (2019, April). A Classification and Data Visualization Tool Applied to Human Migration Analysis. In 2019 Sixth International Conference on eDemocracy & eGovernment (ICEDEG) (pp. 256-261). IEEE.

Donitsa-Schmidt, S. (2017). Social, educational, and linguistic models of immigrants' integration into society. In Beyond Bystanders (pp. 157-167). Brill Sense.

Dunwell, I., Petridis, P., Lameras, P., Hendrix, M., Doukianou, S., & Gaved, M. (2014). Assessing the Reach and Impact of Game-Based Learning Approaches to Cultural Competency. In Proceedings of the 2nd International Workshop on Intelligent Digital Games for Empowerment and Inclusion (IUI2014), arxiv. org, in print.

Duque, A. F., Wolf, G., & Moon, K. R. (2019, October). Visualizing High Dimensional Dynamical Processes. In 2019 IEEE 29th International Workshop on Machine Learning for Signal Processing (MLSP) (pp. 1-6). IEEE.

El Beheiry, M., Doutreligne, S., Caporal, C., Ostertag, C., Dahan, M., & Masson, J. B. (2019). Virtual reality: beyond visualization. Journal of molecular biology, 431(7), 1315-1321.

Esuli, A., & Sebastiani, F. (2006). Sentiwordnet: A publicly available lexical resource for opinion mining. In LREC (Vol. 6, pp. 417-422).

Fish, M. T., Russoniello, C. V., & O'Brien, K. (2018). Zombies vs. Anxiety: An Augmentation Study of Prescribed Video Game Play Compared to Medication in Reducing Anxiety Symptoms. Simulation & Gaming, 49(5), 553–566.

Flanagin, A. J., & Waldeck, J. H. (2004). Technology use and organizational newcomer socialization. The Journal of Business Communication (1973), 41(2), 137-165.

Fortunati, L., Pertierra, R., & Vincent, J. (Eds.). (2013). Migration, diaspora and information technology in global societies. Routledge.

Frank, A. (2013). . The wounded story teller, 2nd ednChicago, IL: The University of Chicago Press.

Gamage, V., & Ennis, C. (2018). Examining the effects of a virtual character on learning and engagement in serious games. Proceedings of the 11th Annual International Conference on Motion, Interaction, and Games - MIG '18, 1–9.

Ganguli, I., Kahn, S., & MacGarvie, M. (Eds.). (2020). The Roles of Immigrants and Foreign Students in US Science, Innovation, and Entrepreneurship. University of Chicago Press.

Garang, R. (2012). Integration and settlement : the experiences and expectations of African immigrants and refugees. Winnepeg Harvest Inc., 2012

Gaver, W. W. (1991). Technology affordances. In Proceedings of the SIGCHI conference on Human factors in computing systems (pp. 79-84).

Gebert, M. (2016). Understanding the Gap between National Labor Regulations and Restrictions vs. Crowdsourcing as a Possibility to Integrate Refugees Fast into Meaningful Labor. Crowdsourcing as a Possibility to Integrate Refugees Fast into Meaningful Labor (January 2, 2016).

Glassner, A. (2017). Interactive storytelling: Techniques for 21st century fiction. CRC Press.

Godfrey, P., Gryz, J., & Lasek, P. (2016). Interactive visualization of large data sets. IEEE Transactions on Knowledge and Data Engineering, 28(8), 2142-2157.

Goodfellow, I., Bengio, Y., Courville, A., & Bengio, Y. (2016). Deep learning (Vol. 1). Cambridge: MIT press.

Gros, B., & Maniega, D. (2018). Co-Designing Apps to Support the Learning Needs of Migrants in Barcelona (Spain). Journal of Interactive Learning Research, 29(3), 465-480.

Gualda, E., & Rebollo, C. (2016). The refugee crisis on Twitter: A diversity of discourses at a European crossroads. Journal of Spatial and Organizational Dynamics, 4(3), 199-212.

Gubrium, A. (2009). Digital storytelling: An emergent method for health promotion research and practice. Health Promotion Practice, 10(2), 186–191.

Gulanowski, D. (2018). The role of online discussion forums in newcomers' labour market integration in Canada. PhD Thesis, Carleton University

Haan, M., McDonald, J.T. (2018). Migration and immigration: recent advances using linked administrative data. J Pop Research 35, 319–324.

Haidle, M. N. (2010). Working-memory capacity and the evolution of modern cognitive potential: implications from animal and early human tool use. Current Anthropology, 51(S1), S149-S166.

Hamilton, M., & Weiss, M. (2005). Children tell stories: Teaching and using storytelling in the classroom. Richard C Owen Pub.

Hannah, J. (2008). The role of education and training in the empowerment and inclusion of migrants and refugees. In Comparative and Global Pedagogies (pp. 33-48). Springer, Dordrecht.

Herrera, F., Bailenson, J., Weisz, E., Ogle, E., & Zaki, J. (2018). Building long-term empathy: A large-scale comparison of traditional and virtual reality perspective-taking. PloS one, 13(10), e0204494.

Houle, R. and Schellenberg, G. New Immigrants' Assessments of Their Life in Canada. 2010

House, R. J., Rousseau, D. M., & Thomas-Hunt, M. (1995). The meso paradigm: a framework for the integration of micro and macro organizational behavior. Research in Organizational Behavior, 17, 71−114.

Hübler, M. (2016). Does migration support technology diffusion in developing countries?. World Development, 83, 148-162.

Hudon, T. Women in Canada : A Gender-based Statistical Report Immigrant Women. Women in Canada 7th Editio, 89 (2015), 1–39

Hwang, G. J., & Wu, P. H. (2012). Advancements and trends in digital game-based learning research: a review of publications in selected journals from 2001 to 2010. British Journal of Educational Technology, 43(1), E6-E10.

Indurkhya, N., & Damerau, F. J. (Eds.). (2010). Handbook of natural language processing (Vol. 2). CRC Press.

IOM (2018). GLobal Migration Indocators.

IOM (2019). Migration Glossary, International Organization for Migration. https://www.iom.int/glossary-migration-2019

Johnson, L., & Kendrick, M. (2017). "Impossible is nothing": Expressing difficult knowledge through digital storytelling. Journal of Adolescent & Adult Literacy, 60(6), 667-675.

Keating, B. (2017). Changing Hearts and Minds, using Virtual Reality to Improve Empathy towards Refugees. Prehospital and Disaster Medicine, 32(S1), S86-S86.

Kirkland, K., & Williamson, B. (2010, October). Play-school: Linking culture and curriculum through games-based learning in schools. In Proceedings for the 4th European Conference on Games Based Learning (pp. 168-176).

Klein, K., Sommer, B., Nim, H. T., Flack, A., Safi, K., Nagy, M., ... & Quetting, M. (2019). Fly with the flock: immersive solutions for animal movement visualization and analytics. Journal of the Royal Society Interface, 16(153), 20180794.

Kniffin, K. M., Narayanan, J., Anseel, F., Antonakis, J., Ashford, S. P., Bakker, A. B., ... & Vugt, M. V. (2021). COVID-19 and the workplace: Implications, issues, and insights for future research and action. American Psychologist, 76(1), 63.

Kopala, M., Esquivel, G., & Baptiste, L. (1994). Counseling Approaches for Immigrant Children: Facilitating the Acculturative Process. The School Counselor, 41(5), 352–359

Kors, M. J., Ferri, G., Van Der Spek, E. D., Ketel, C., & Schouten, B. A. (2016, October). A breathtaking journey. On the design of an empathy-arousing mixed-reality game. In Proceedings of the 2016 Annual Symposium on Computer-Human Interaction in Play (pp. 91-104).

Lamber, R., Aigner, A., Kappel, K., Baranyi, R., & Grechenig, T. (2018, November). A Hands-on Technology Mix for a European Refugee Management System serving Migrants and Institutions. In 2018 IEEE 9th International Conference on Software Engineering and Service Science (ICSESS) (pp. 26-31). IEEE.

McBrien, J L. (2005). Educational Needs and Barriers for Refugee Students in the United States: a Review of the Literature. Review of Educational Research, 75(3): 329-364.

McDonough, S., & Colucci, E. (2019). People of immigrant and refugee background sharing experiences of mental health recovery: reflections and recommendations on using digital storytelling. Visual Communication, 1470357218820651.

McGregor, E., & Siegel, M. (2013). Social media and migration research (Working Paper No. 2013- 068). Maastricht: United Nations University-Maastricht Economic and Social Research Institute on Innovation and Technology (MERIT).

McLean, K. C., & Syed, M. (2015). Personal, master, and alternative narratives: An integrative framework for understanding identity development in context. Human Development, 58(6), 318–349.

Millais, P., Jones, S. L., & Kelly, R. (2018, April). Exploring data in virtual reality: Comparisons with 2d data visualizations. In Extended Abstracts of the 2018 CHI Conference on Human Factors in Computing Systems (pp. 1-6).Weiskopf, D. (2007). GPU-based interactive visualization techniques. Berlin: Springer.

Miller, K. E. & Billings, D. L. (1994). Playing to grow: A primary mental health intervention with Guatemalan refugee children. American Journal of Orthopsychiatry, Vol 64(3), Jul 1994, 346-356

Miller, S. (2018). Monitoring Migrants in the Digital Age: Using Twitter to Analyze Social Media Surveillance. Colo. Tech. LJ, 17, 395.

Mislove, A., Sune Lehmann, Yong-Yeol Ahn, Jukka-Pekka Onnela, J.Neils Rosenquinst. (2011) Understanding demographics of Twitter Users. In Proceedings of the Fifth International AAAI Conference on Weblogs and Social Media.

Moffitt, U. E., Nardon, L., & Zhang, H. (2019). Becoming Canadian: Immigrant narratives of professional attainment. International Journal of Intercultural Relations.

Mortara, M., Catalano, C. E., Bellotti, F., Fiucci, G., Houry-Panchetti, M., & Petridis, P. (2014). Learning cultural heritage by serious games. Journal of Cultural Heritage, 15(3), 318-325.

Murray, J. (1998). Hamlet on the Holodeck. MIT press.

Musolesi, M. (2014). Big mobile data mining: Good or evil?. IEEE Internet Computing, 18(1), 78-81.

Nakhaie, M. R. (2018). Service needs of immigrants and refugees. Journal of International Migration and Integration, 19(1), 143-160.

Nardon, L., Aten, K., & Gulanowski, D. (2015). Expatriate adjustment in the digital age: The co-creation of online social support resources through blogging. International journal of intercultural relations, 47, 41-55.

Ohler, Jason B. (2013). Digital storytelling in the classroom: New media pathways to literacy, learning, and creativity. Corwin Press

Olsen, L., & Dowell, C. (1989). Bridges: Promising Programs for the Education of Immigrant Children.

Pannocchia, D., Bayerl, P. S., & Hough, K. L. (2020). The digital mediation of migration: a qualitative thematic synthesis. In: NUNES, Miguel Baptista, ISAÍAS, Pedro, POWELL, Philip and BONTCHEV, Boyan, (eds.) Proceedings of the 13th IADIS International Conference Information Systems 2020.IADIS Press, 145-150

Pennacchiotti, M. and Ana-Maria Popescu (2011). A Machine Learning Approach to Twitter User Classification. In Proceedings of the Fifth International AAAI Conference on Weblogs and Social Media, 2011

Pianta, E., Bentivogli, L., & Girardi, C. (2002). MultiWordNet: developing an aligned multilingual database. In First international conference on global WordNet (pp. 293-302).

Puiu, D., Barnaghi, P., Tönjes, R., Kümper, D., Ali, M. I., Mileo, A., ... & Gao, F. (2016). Citypulse: Large scale data analytics framework for smart cities. IEEE Access, 4, 1086-1108.

Quach, Q., & Jenny, B. (2020). Immersive visualization with bar graphics. Cartography and Geographic Information Science, 1-10.

Rao, D.; D., Y.; Shreevats, A.; and Gupta, M. (2010). Classifying Latent User Attributes in Twitter. In Proceedings of SMUC-10, 710–718, 2010

Rapson, N. (2020). Facilitation of immigrant integration in Canada: Storytelling for educational understanding (Doctoral dissertation, Concordia University).

Reis, A., Nunes, D., Aguiar, H., Dias, H., Barbosa, R., Figueira, A., ... & Boavida, F. (2016). Tech4SocialChange: crowdsourcing to bring migrants experiences to the academics.

Risam, R. (2019). Beyond the Migrant "Problem": Visualizing Global Migration. Television & New Media, 20(6), 566-580.

Roda, F., Zanni-Merk, C., & CIFASIS, C. U. A. (2016, June). An Intelligent Data Analysis Framework for Supporting Perception of Geospatial Phenomena. In FOIS (pp. 95-108).

Rousseau, C., & Heusch, N. (2000). The trip: A creative expression project for refugee and immigrant children. Art Therapy, 17(1), 31-40

Rousseau, C., Drapeau, A., Lacroix, L., Bagilishya, D., & Heusch, N. (2005). Evaluation of a classroom program of creative expression workshops for refugee and immigrant children. Journal of Child Psychology and Psychiatry, 46(2), 180-185

Rozmi, M. D. A. B., Thirunavukkarasu, G. S., Jamei, E., Seyedmahmoudian, M., Mekhilef, S., Stojcevski, A., & Horan, B. (2019). Role of immersive visualization tools in renewable energy system development. Renewable and Sustainable Energy Reviews, 115, 109363.

Sankaranarayanan, H. B., Agarwal, G., & Rathod, V. (2016, October). An exploratory data analysis of airport wait times using big data visualisation techniques. In 2016 International Conference on Computation System and Information Technology for Sustainable Solutions (CSITSS) (pp. 324-329). IEEE.

Sano, A., & Picard, R. W. (2013). Stress Recognition Using Wearable Sensors and Mobile Phones. 2013 Humaine Association Conference on Affective Computing and Intelligent Interaction, 671–676.

Saravanan, M. (2017, July). Determining ethnicity of immigrants using Twitter data. In Proceedings of the 4th Multidisciplinary International Social Networks Conference (pp. 1-10).

Saridaki, M., & Meimaris, M. (2018). Digital Storytelling for the empowerment of people with intellectual disabilities. In Proceedings of the 8th International Conference on Software Development and Technologies for Enhancing Accessibility and Fighting Info-exclusion (pp. 161-164).

Scavarelli, A., Arya, A., & Teather, R. J. (2020). Virtual reality and augmented reality in social learning spaces: a literature review. VIRTUAL REALITY.

Shen, Y., He, X., Gao, J., Deng, L., & Mesnil, G. (2014, April). Learning semantic representations using convolutional neural networks for web search. In Proceedings of the 23rd international conference on world wide web (pp. 373-374).

Shilton, K. (2009). Four billion little brothers? Privacy, mobile phones, and ubiquitous data collection. Communications of the ACM, 52(11), 48-53.

SPCO (2010). Immigrant Children, Youth and Families: A Qualitative Analysis of the Challenges of Integration, Social Planning Council of Ottawa

Stafford, S. P. (2009). Data, information, knowledge, and wisdom. Knowledge Management, Organizational Intelligence And Learning, And Complexity 3:179.

Stewart, M., Anderson, J., Beiser, M., et al. Multicultural meanings of social support among immigrants and refugees. International Migration 46, 3 (2008), 123–159

Su, S., Perry, V., An, M., Bravo, L., & Dasari, V. (2019, May). HPC enabled immersive and non-immersive visualization of large scale scientific data. In Disruptive Technologies in Information Sciences II (Vol. 11013, p. 110130P). International Society for Optics and Photonics.

Syed, J. (2008). Employment prospects for skilled migrants: A relational perspective. Human Resource Management Review, 18(1), 28-45.

Tadeja, S. K., Seshadri, P., & Kristensson, P. O. (2020). AeroVR: An immersive visualisation system for aerospace design and digital twinning in virtual reality. The Aeronautical Journal, 1-21.

Taeihagh, A. (2017). Crowdsourcing, sharing economies and development. Journal of Developing Societies, 33(2), 191-222.

Tanenbaum, J. (2014). Design fictional interactions: why HCI should care about stories. interactions, 21(5), 22-23.

Tattersall, I. (2009). Human origins: Out of Africa, Proceedings of the National Academy of Sciences Sep 2009, 106 (38) 16018-16021

Tominski, C., & Schumann, H. (2020). Interactive Visual Data Analysis. CRC Press.

Turk, M. (2014). Multimodal interaction: A review. Pattern Recognition Letters, 36, 189-195.

Underhill, E., Huang, S., Yi, S., & Rimmer, M. (2019). USING SOCIAL MEDIA TO IMPROVE TEMPORARY MIGRANT WORKERS'ACCESS TO INFORMATION ABOUT THEIR EMPLOYMENT RIGHTS. The Journal of Australian Political Economy, (84), 147-174.

Van Neste-Gottignies, A. (2019). 1.2. 3 Communication towards/with migrants. Organizer, 46.

Von Neumann, J. (1945). First draft of a report on the edvac, contract no. w-670-ord-402 moore school of electrical engineering, univ. of penn., philadelphia..

Wang, J., He, X., Wang, Z., Wu, J., Yuan, N. J., Xie, X., & Xiong, Z. (2018, April). CD-CNN: a partially supervised cross-domain deep learning model for urban resident recognition. In Thirty-Second AAAI Conference on Artificial Intelligence.

Wei, Y. (2020). The roles of digital storytelling in the construction of identity and belonging among young Chinese immigrants in Australia (Doctoral dissertation, The University of Melbourne).

Wieland, M. L., Njeru, J. W., Hanza, M. M., Boehm, D. H., Singh, D., Yawn, B. P., ... & Goodson, M. (2017). Pilot feasibility study of a digital storytelling intervention for immigrant and refugee adults with diabetes. The Diabetes Educator, 43(4), 349-359.

Wihbey, J. P., Jackson, S. J., Cruz, P. M., & Welles, B. F. (2020). 22. Visualizing diversity: Data deficiencies and semiotic strategies. Data Visualization in Society, 369.

Wister, P. Pancardo, F. D. Acosta, J. A. Hernandez-Nolasco and E. S. Delacruz, Experiences with RFID technology to manage information during natural disasters, 2015.

Woolcock, M. (2001). The place of social capital in understanding social and economic outcomes. Canadian Journal of Policy Research 2, 1 (2001), 1–35

Wright, P., & McCarthy, J. (2008). Empathy and experience in HCI. In Proceedings of the SIGCHI conference on human factors in computing systems (pp. 637-646).

Wrigley, H. (2007). Beyond the life boat: Improving language, citizenship and training services for immigrants and refugees. Toward defining and improving quality in adult basic education: Issues and challenges, 221-239.

www.quantcast.com

Yang, Ya-Ting C., and Wan-Chi I. Wu. (2012). Digital storytelling for enhancing student academic achievement, critical thinking, and learning motivation: A year-long experimental study. Computers & Education 59.2 : 339-352.

Yao, S., Jennings, S., & Dai, Z. (2019). Using Social Media to Increase Psychological Well-Being of Chinese Immigrants: A Case Study of MySunnysky--an Online Counseling and Therapy Service. China Media Research, 15(3).

Zijlstra, J., & Liempt, I. V. (2017). Smart (phone) travelling: Understanding the use and impact of mobile technology on irregular migration journeys. International Journal of Migration and Border Studies, 3(2-3), 174-191.